\documentclass[namedreferences,hyperref,optionalrh,solaromanenum]{spr-sola}

\usepackage{graphicx}                    
\usepackage{color}                       

\chardef\us=`\_

\begin{document}

\begin{frontmatter}

\title{Estimation of the lifetime of slow-decaying unipolar active regions in the framework of the turbulent erosion model}

\author[addressref={},corref,email={}]{\inits{}\fnm{}\lnm{}\orcid{}}

\author[addressref={aff1},corref,email={plotnikov.andrey.alex@yandex.ru}]{\inits{A.A.}\fnm{Andrei}~\lnm{Plotnikov}}
\author[addressref=aff1]{\inits{V.I.}\fnm{Valentina}~\lnm{Abramenko}}
\author[addressref=aff1]{\inits{A.S.}\fnm{Alexander}~\lnm{Kutsenko}}
\address[id={aff1}]{Crimean Astrophysical Observatory, Russian Academy of Sciences, Nauchny, 298409, Russia}

%
\runningtitle{Lifetime of slow-decaying ARs}


\begin{abstract}
We explore properties and behavior of slow-decaying unipolar sunspot groups in the framework of the turbulent erosion model suggested by \citet{PetrovayMorenoInsertis1997}. The basic concept of the model is the suppression of a turbulent diffusivity inside a magnetic flux tube by strong magnetic fields. As a result, the outer turbulent plasma detaches magnetic features primarily from the external border of a magnetic flux tube. The radius of the tube exhibits inward progression at a constant rate. The model predicts older sunspots to decay slower and it seems to be very promising to explain the slow decay of long-living unipolar sunspot groups. By analyzing the magnetic structure associated with a sunspot, we did reveal a gradual decrease of the magnetic structure radius at a constant rate implying the validity of the model. However, in some cases the derived velocity of the radius decrease was too low: our calculations provided implausibly high estimations for the lifetime and maximal area of such sunspots. We discuss possible additional mechanisms affecting the decay rate of such peculiar sunspots.
\end{abstract}

%

\end{frontmatter}

\section{Introduction}
The lifetime of a solar active region (AR) seen as a sunspot group in continuum intensity images could be divided into two parts: the phase of growth (or emergence) and the phase of fade (or decay). The process of the sunspot group decay was studied over the decades and various explanations were suggested.
\citet{Gokhale1975} proposed a ``self-similar sunspot'' model, where the decay is provided by Ohmic dissipation of electric currents and, therefore, of the magnetic flux and sunspot's area. Later, \citet{Meyer1974} and \citet{Krause1975} suggested the ``turbulent diffusion'' model in which the magnetic flux tubes are dispersed by the diffusion motions of plasma.

\citet{PetrovayMorenoInsertis1997} proposed a ``turbulent erosion'' model (hereinafter referred to as PMI model) that is widely acknowledged nowadays \citep[\textit{e.g.},][]{Dacie, Litvinenko, Xu}. The main idea of PMI model is a suggestion that the turbulent diffusivity is suppressed significantly with the increase of the magnetic field. Under this assumption, diffusive motions seem to be ``frozen'' in the internal parts of strong magnetic structures and take no part in the process of decay. This might result in a situation when the turbulent motions of the outer plasma flows ``gnaw'' the border of the magnetic structure. A parabolic law of the decay follows from  PMI model, which is consistent with statistical studies \citep{PetrovayDrielGesztelyi, Murakozy} in contrast to the ``self-similar sunspot'' and ``turbulent diffusion'' models.

\citet{PetrovayMorenoInsertis1997} describe the decay process (as a decrease of the sunspot area) with a parabolic law:
\begin{equation} \label{parabolic_law}
	A = A_0 - 2\sqrt{\pi A_0} w t + \pi w^2 {t}^2,
\end{equation}
where $A_0$ denotes the maximal area of a sunspot, and $t$ is the current time (the decay starts at the moment $t = 0$). The parameter $w$ denotes the speed of the magnetic diffusivity front, in other words, the speed of the decrease of the sunspot radius. This value is assumed (in PMI model) to be time-independent along the decay phase, but to be different for different sunspots.    

Using Equation~\ref{parabolic_law}, the time derivative of the area can be written as:
\begin{equation} \label{decay_rate_parabolic}
    \dot{A} = -2\sqrt{\pi A_0} w + 2 \pi  w^2 {t}.
\label{A_dot}    
\end{equation}
At time 
\begin{equation}
t = T = \frac{1}{w} \sqrt{\frac{A_0}{\pi}}
\label{eq:lifetime}
\end{equation}
quantities $\dot{A}$ and $A$ become zero, and the sunspot disappears. The interval $T$ is the duration of the decay process. Henceforward we will treat the interval $T$ as the ``lifetime'' of a sunspot group keeping in mind that the decay phase is the longest phase in the whole evolution of a sunspot group \citep[\textit{e.g.},][]{UgarteUrra2015}.

Recently \citet{Plotnikov} performed a statistical study of the magnetic flux decay in hundreds of ARs. The authors revealed a cluster of slow-decaying ARs exhibiting significantly lower magnetic flux decay rate than that expected from the common (for all ARs) power-law dependence. All of these ARs were found to be unipolar sunspot groups in continuum intensity images, {\it i.e.,} the sunspot group consisted of a single sunspot related to (as a rule) the leading magnetic polarity. The following polarity exhibited neither sunspots nor pores in continuum intensity images. Throughout the rest of the paper, we consider exclusively such unipolar $\alpha$-Hale class sunspot groups, namely, the magnetic structure related to the sunspot itself.

Equation~\ref{A_dot} assumes the decay rate to decrease along the decay phase. Thus, in the framework of PMI model, we can make a conclusion: when we observe two sunspots with similar areas but with different decay rates, we can expect that the fast-decaying one just started its decay while the slow-decaying one was decaying for a long time (\textit{e.g.}, more than one solar rotation). Keeping in mind that the area of a sunspot group is proportional to the total magnetic flux of the corresponding AR, the above mentioned set of slow-decaying ARs in \citet{Plotnikov} can help us to reveal whether this conclusion is justified by observational data. 

In this work, we estimate the maximal areas of some slow-decaying sunspots and their lifetimes in the framework of PMI model to check whether their behaviour can be explained by the mechanism of the turbulent erosion. 

\section{Formalism}

PMI model considers the decay of a magnetic flux tube in a turbulent medium. According to the model, the speed of decrease of the tube radius, $w$, can be expressed as:
\begin{equation}
w \sim \frac{B_e}{B_0}\frac{\nu_0}{r_0},
\label{eq_speed}   
\end{equation}
where $B_e$ denotes the cutoff magnetic field (a model parameter), $B_0$ is the peak magnetic field in the center of the tube, $\nu_0$ is the quiet-Sun magnetic diffusivity, and $r_0$ is the maximal radius of the tube. Note that, similar to $w$, the value of $B_0$ is assumed to be constant during the entire decay phase for the selected magnetic flux tube although it might differ for different tubes.

Transforming, we get:
$$
{r_0} \sim \frac{B_e}{B_0}\frac{\nu_0}{w}
$$
So, the value of $r_0$ can be estimated from the known magnitude of $w$. $B_e$ and $\nu_0$ are the model parameters, and can not be derived from the observations. However, they will disappear when we undertake a comparison of two independent magnetic elements (sunspots) that are assumed to be the cross-sections of the magnetic tubes in the photosphere.

Indeed, suppose we have two magnetic elements with the parameters ($B_{01}$, $w_{1}$) and ($B_{02}$, $w_{2}$), respectively. 
Then,
$$
\frac{r_{02}}{r_{01}} = \frac{w_1}{w_2} \frac{B_{01}}{B_{02}},
$$

or,

$$
r_{02} = r_{01} \frac{w_1}{w_2} \frac{B_{01}}{B_{02}}.
$$

Converting to areas, we have:
\begin{equation}
A_{02} = A_{01} \left(\frac{w_1}{w_2} \frac{B_{01}}{B_{02}}\right)^2.
\label{eq:area_ref}
\end{equation}

Sometimes, the magnitude of $A_{0}$ is unknown (when the start moment of the decay phase was not observed). In this case, we can estimate only the lower boundary of $A_{0}$. For an arbitrary time $t$ during the decay (as well as for the time of the start of the observations), we have:
$$
A(t) \leq A_{0}.
$$
Thus:
\begin{equation}
A_{02} \geq A_{1}(t) \left(\frac{w_1}{w_2} \frac{B_{01}}{B_{02}}\right)^2.
\label{eq:area_lower}
\end{equation}

Now, the lifetime of the magnetic element can be estimated using Equation~\ref{eq:lifetime}.

As it was mentioned above, $w$ is the speed of decrease of the magnetic element's radius. Hence:  
\begin{equation}
    \dot{A} = -w P,
\label{eq:derivative}
\end{equation}
where $A$ and $P$ are the area and and the perimeter of the magnetic element, respectively. In this work, the areas and perimeters of magnetic element were derived using the magnetic field maps. In contrast to previous works, dealing primarily with continuum intensity images, nowadays solar magnetographic observations allow us to analyze the magnetic flux bundle decay, for which PMI model was developed. As we mentioned above, the magnetic elements are associated with the single sunspot in unipolar sunspot groups. The details of data processing are presented in the next section. 

\section{Data and Methods}

We used data acquired by the \textit{Helioseismic and Magnetic Imager} on board the \textit{Solar Dynamics Observatory} \citep{Scherrer2012}. SHARP CEA \citep{Bobra2014} are the patches of the magnetic field vector, re-projected to the cylindrical equal-area coordinates. This coordinate system allows us to compensate for the effects of the projection and to derive more accurate estimates for the magnetic flux and the perimeter values. We used patches with the cadence of 1 hour.   

\begin{figure*}
    \centering
    \includegraphics[width = 1\columnwidth]{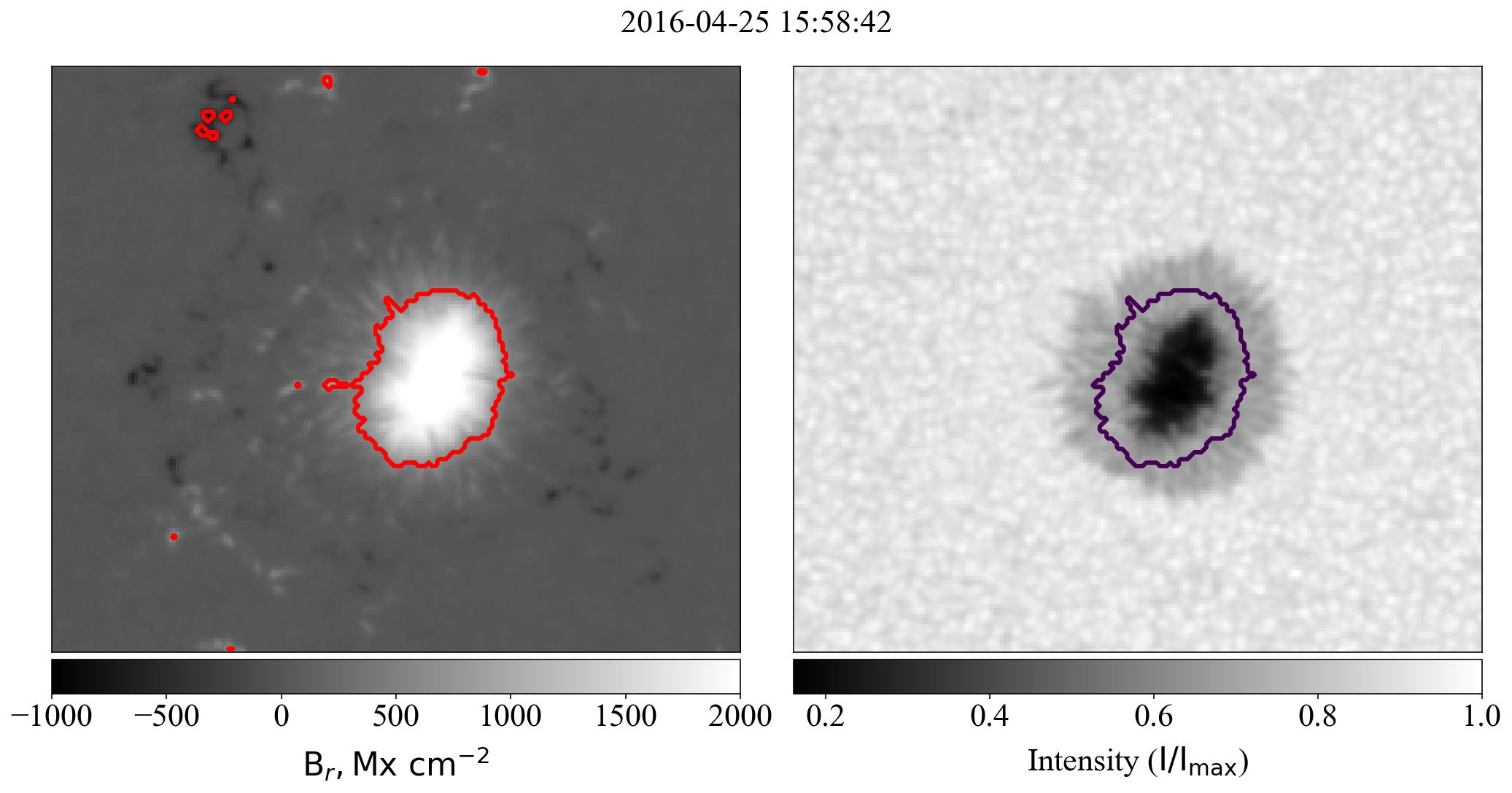}
    \caption{Radial magnetic field map and continuum intensity image of NOAA AR 12533. Red contours in the left panel show the masked magnetic elements with the pixel magnetic flux density exceeding the 600~Mx~cm$^{-2}$ threshold. The purple contour in the right panel demonstrates the largest magnetic element selected for further analysis. Note that the magnetic element's border coincide with neither the umbra nor the penumbra of the sunspot.}
    \label{fig:contours}
\end{figure*}

We define the perimeter $P$ of a magnetic element as a length of an isoline along the cutoff strength of the radial magnetic field (Figure~\ref{fig:contours}, left panel). The area is defined as an area inside the isoline. The cutoff threshold was set to 600~Mx~cm$^{-2}$, according to \citet{Norton2017}. This value is related to the error level of the transverse field derived from SDO/HMI, used for calculation of the radial field (note that \citet{Norton2017} used the threshold of 575~Mx~cm$^{-2}$). All magnetograms were masked to omit pixels of the magnetic field lower than the cutoff threshold. 
The derived masks consisted a set of isolated singly connected magnetic elements shown with the red contours in the left panel of Figure~\ref{fig:contours}. The element with the largest area (containing the largest number of pixels) represented the unipolar spot itself and was chosen as the magnetic element to be analyzed (shown with the purple contour in the right panel of Figure~\ref{fig:contours}). 

It should be noted that the described method is suitable only for unipolar sunspot groups, where the prevailing magnetic feature can be unequivocally associated with the single sunspot itself. Additional difficulties may occur with sunspots that have light bridges or break into parts during the observations. Consequently, we chose manually a set of 6 unipolar sunspot groups without such features.  

The perimeter of the magnetic element was calculated using the four-neighbours method \citep{Benkrid}. The area was calculated as the number of pixels within the element.

\begin{figure*}
    \centering
    \includegraphics[width = 1\columnwidth]{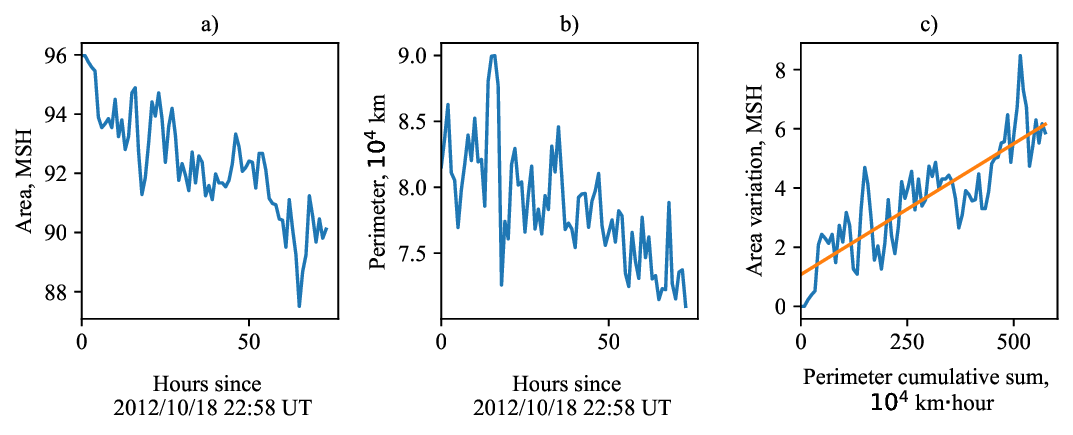}
    \caption{Time variations of the area (panel $\textbf{a}$) and the perimeter (panel $\textbf{b}$)  for the sunspot-associated magnetic element in} AR NOAA 11591. Panel \textbf{c} - the linear fit (red line) to the area-vs-perimeter plot to calculate the magnitude of the speed of the sunspot radius' decrease, $w$, as a slope of the fit.  
    \label{fig:plots}
\end{figure*}

For each of the considered magnetic elements, we calculated the area and perimeter variations in time. An example of the area \textit{versus} time and perimeter \textit{versus} time curves for the sunspot-associated magnetic element in NOAA AR 11591 is shown in the left and middle panels of Figure~\ref{fig:plots}, respectively.

Noise in the source data will make the direct calculations of $w$ from Equation~\ref{eq:derivative} inaccurate. To mitigate the noise issue, we consider the integral form of Equation~\ref{eq:derivative}:

\begin{equation}
    A(t) - A(t_0) = -w \int_{t_0}^{t}{P \,dt}.
\end{equation}

In terms of discrete variables, the integral turns into a cumulative sum (time intervals $\Delta t$ can be included into the $w$ value in case of equidistant intervals): 
\begin{equation}
    A_n - A_0 = -w \sum_{0}^{n}{P_i}.
\end{equation}

Calculations for consecutive time intervals $t_i$ gave us two sets of values: variations of the magnetic element's area and variations of the cumulative sum of the perimeters. Then the magnitude of $w$ was found as a slope of the linear approximation between these sets (Figure~\ref{fig:plots}c). The area decay rate, $\dot{\textrm{A}}$, was calculated as the slope of the linear fit of the area versus time function (Figure~\ref{fig:plots}a).

\section{Results}
Table \ref{tbl:common} shows the values of the parameters calculated for the set of 6 unipolar sunspot groups. The groups are arranged by the ascending order of the area decay rate $\dot{\textrm{A}}$. The maximal observed area listed   in the table is the maximal area measured during the observational interval. 

Note that the area decay rate changes (from the lowest to the largest values) by 5 times, the speed of the radius' decrease, $w$, changes by 3.5 times for the selected sunspot groups. At the same time, the area and the perimeter change by 2.3 and 1.8 times, respectively. So, while the area and perimeter of the selected unipolar sunspot groups might be rather close to each other, their decay rate $\dot{\textrm{A}}$ and their speed of the radius' decrease, $w$, can differ more significantly.

Using Equation~\ref{eq:area_lower}, we can estimate the lower boundary of the maximal area of a magnetic element at the maximum development (just after the emergence halts). This was done using a pair of sunspot groups, one of them being the reference. The $B_0$ value in Equation~\ref{eq:area_lower} was set as the peak magnetic flux density in the magnetic element's center at the beginning of observations. Recall that $B_0$ is assumed to be a constant during the decay. 
Then the lifetime was derived from Equation~\ref{eq:lifetime}. The derived peak area and lifetime through the rest of the text will be considered as the expected from the model estimetes.

Table \ref{tbl:lifetimes} contains the values of the expected maximal area and the lifetime estimated using Equations~\ref{eq:area_lower}~and~\ref{eq:lifetime}. The values were calculated with three different sunspots as a reference, and placed in the separate columns. Using of three reference sunspots allows us to perform three independent estimates of the expected area and lifetime, which turned to be correlated well to each other (for example, for the sunspot in NOAA 12533 the lifetime from three references is obtained as 503, 484, and 432 days). 

For the three upper slow-decaying sunspot groups in Table \ref{tbl:lifetimes}, the calculations gave no realistic estimates for the lifetime. For example, the sunspot in NOAA 12246 should have the lifetime longer than 2 years. Moreover, these values are only the lower boundary. In the framework of PMI model, the low decay rate is a result of an old age of a sunspot group, but not so old as 2 years. The estimated values of the lifetime contradict observations: none of the sunspot groups considered in this work existed for more than two solar rotations. 

\begin{table}
\caption{Parameters derived from observations.}
\label{tbl:common}
\begin{tabular}{ccccccc}     
\hline
NOAA & Dates  & Max. observed & $\dot{\textrm{A}}$, & $w$ , & Max. perimeter, & $B_0$,\\
    number & of observations (UT)  &area, MSH & $10^{-2}$ MSH $\cdot h^{-1}$ & km $\cdot\ h^{-1}$ & $10^4$~km & G \\
\hline
12246 & 2014/12/23 05:58 --  & 66.94 & 2.56 & 1.44 & 6.65 & 2846\\
 & 2015/01/01 08:58        &       &     &      &     & \\ 
12533 & 2016/04/21 19:58 --  & 85.92 & 3.66 & 1.76 & 8.08 & 3138\\
 & 2016/04/30 17:58        &       &     &      &     & \\ 
11591 & 2012/10/18 22:58 --  & 95.95 & 6.97 & 2.69 & 9.00 & 3467\\
 & 2012/10/21 23:58        &       &     &      &      &\\ 
11649 & 2013/01/05 20:58 --  & 66.68 & 8.27 & 4.86 & 8.14 & 2580\\
 & 2013/01/13 21:58       &       &     &      &      &\\ 
11809 & 2013/08/04 18:58 --  & 41.27 & 9.68 & 6.96 & 5.31 & 2568\\
 & 2013/08/10 14:58        &       &     &      &      &\\ 
12348 & 2015/05/15 08:58 --  & 77.80 & 12.61 & 5.13 & 9.70 & 2635\\
 & 2015/05/19 01:58      &       &     &      &      &\\ 

\end{tabular}
\end{table}

\begin{table}
\caption{Parameters derived from PMI model}
\label{tbl:lifetimes}
\begin{tabular}{ccccccc}     
\hline
~ & \multicolumn{2}{c}{Ref. AR 12348}  & \multicolumn{2}{c}{Ref. AR 11809} &  \multicolumn{2}{c}{Ref. AR 11649}\\
NOAA & $A^{exp}_0$, MSH  & $T^{exp}$, days & $A^{exp}_0$, MSH  & $T^{exp}$, days & $A^{exp}_0$, MSH & $T^{exp}$, days\\
\hline
12246 & 846 & 829 & 785 & 798 & 624 & 711 \\
12533 & 466 & 503 & 432 & 484 & 344 & 432 \\
11591 & 163 & 195 & 152 & 188 & 121 & 167 \\
11649 & 90 & 80 & 84 & 77 & -~ & -~ \\
11809 & 44 & 39 & -~ & -~ & 33 & 34 \\
12348 & -~ & -~ & 72 & 68 & 57 & 61 \\

\end{tabular}
\end{table}

\section{Discussion and Conclusions}
In this work, we explore properties and behaviour of slow-decaying sunspot in unipolar sunspot groups. The aim was to find whether the turbulent erosion can explain the very low decay rate of certain sunspots surviving up to 2--3 solar rotations. The turbulent erosion model of the sunspot decay suggested by \citet{PetrovayMorenoInsertis1997}, PMI model, predicts that the older the sunspot, the slower it decays. The model seems to be very promising to explain the phenomenon. However, we found that in certain cases the model provides implausibly high estimations for the lifetime of such sunspots.

Statistical works studying a relationship between the sunspot group's maximal area and its lifetime \citep[also known as Gnevyshev-Waldmeier law, \textit{e.g.}][]{Bradshaw, Nagovitsyn} give lower estimations for the sunspot group lifetimes than the present work does. This discrepancy might be explained by the following considerations. The above cited works explored white-light images, whereas we considered here the magnetic elements corresponding to sunspots. It is well known that the later survive longer. Second, the statistical studies mentioned above are based on hundreds of very different sunspot groups of all possible configurations and evolution tracks. At the same time, in the present study, only a few sunspot groups of a specific morphology are explored. All of them are unipolar sunspot groups and a part of them is associated with a subset of slow-decaying ARs \citep{Plotnikov}. Note that the present study does not include the cases when a sunspot breaks into parts, or a light bridge in the sunspot appears, which can dramatically change the course of the ``quiet'' erosion by crushing the magnetic element and increasing the perimeter.

PMI model uses granular value of the turbulent diffusivity ($\nu_0 \sim 1000~\textrm{km}^2$) for estimations of the speed of the diffusivity front.  For the threshold magnetic field $B_e$ of 400~Mx~cm$^{-2}$ and typical sunspot values of the maximal magnetic field $B_0 \sim 3000$~Mx~cm$^{-2}$ and the maximal sunspot radius $r_0 \sim 10^4~$km (corresponding to 100~MSH sunspot's area), we got the speed of the radius' decrease (the speed of the diffusivity front) $w$ of about 50~km$\cdot \textrm{h}^{-1}$. This is approximately one order of magnitude higher than that derived in the present work from observations.

In their original work \citet{PetrovayDrielGesztelyi} tested PMI model by fitting the area decay rate of individual sunspots by a parabolic decay law. The parabolic law approximated the distribution quite well, especially for binned data, although the scatter of the points was large \citep[see Figure~1 in][]{PetrovayDrielGesztelyi}. The sample size in the present study is too small to perform any reliable fitting or statistical tests. Nevertheless, the linear relationship between the area and perimeter of the magnetic element (Figure~\ref{fig:plots}c) implies the inward movement of the current sheet at a constant speed $w$ as the decay proceeds. This point supports the basic idea of PMI model, although the derived value $w$ is too low in some cases.

PMI model also gives a generalization of the turbulent erosion in the presence of radial outflows that are often observed around sunspots \citep[(so-called magnetic moats, \textit{e.g.}][]{Harvey1973, Yurchyshyn2001}. In this case, an additional term in Equation~\ref{eq_speed} appears that may affect the speed of the current sheet:
\begin{equation}
    w \geq \frac{B_e}{B_0} \left( \frac{\nu_0}{r_0} + v_0 \right).
\end{equation}
In other words, the radial outflow might regulate the decay rate. The deficit of the radial outflow velocity, or even inflows, could be one of explanations for the slow-decaying sunspots. Values of $\nu_0 \sim 1000~\textrm{km}^2$ and $r_0 \sim 10^4~$km give a characteristic value of $v_0 \sim 0.1~\textrm{km} \cdot \textrm{s}^{-1}$. 
Variations in the radial velocity can make the decay rate vary by several times. \cite{Kubo2008} reported the outflow velocities of 0.5--1 $\textrm{km} \cdot \textrm{s}^{-1}$, so, the existence of such a deficit looks possible. The inflows are not confirmed by the observations in the photosphere but may be present in the subsurface levels.

\citet{Petrovay1999} performed a thorough analysis of possible reasons for deviations from the mean parabolic law suggested by PMI model. The authors argued that surrounding (plage) magnetic fields might affect strongly the speed of the current sheet \citep[see Figure~6 and sections 4-6 in][]{Petrovay1999}, namely a high value of the background magnetic field strength decreases the speed of the current sheet. This background field might be a pre-existing one (existed before the emergence of the sunspot group), or generated during the decay of the sunspot itself. In the latter case, the ``memory'' about the initial size of the sunspot appears: a large magnetic structure produces more plage fields slowing down the further decay. Randomness and variability of the background magnetic field might result in scatter of the decay rate. Anyway, all of the aforementioned assumptions require further careful research.

%

%

%

%
\begin{acks}
We are grateful to the anonymous referee whose comments helped us to improve the paper significantly. SDO is a mission for NASA’s Living With a Star (LWS) programme. The SDO/HMI data were provided by the Joint Science Operation Center (JSOC). Python programming language with NumPy \citep{numpy}, SciPy \citep{scipy} and SunPy \citep{sunpy} libraries was used for the numerical analysis. All plots were made with using of Matplotlib \citep{matplotlib} library.
\end{acks}

%
%
%
%
%
%
%

%
%
\bibliographystyle{spr-mp-sola}
\bibliography{Lifetime}  
%
%
%
%

\end{document}